%% file: EQW_def-v2.tex
\documentclass[pra,twocolumn,superscriptaddress,groupedaddress,a4paper]{revtex4}

\usepackage{amsmath,amsthm,amsfonts,amssymb,times,graphicx,color}
\usepackage{multirow}
\usepackage{subfigure}
\usepackage{mathtools}
\usepackage{url}
\usepackage{color}
\usepackage{hyperref}

\include{kkmacros}

\def\PA#1{{\textcolor{magenta}{}}}

\begin{document}

\title{The elephant quantum walk}

\author{Giuseppe Di Molfetta}
\email[]{giuseppe.dimolfetta@lis-lab.fr}
\affiliation{Aix-Marseille Universit\'{e}, Universit\'{e} de Toulon, CNRS, LIS, Marseille, France,
Natural computation research group}
\affiliation{and Departamento de F{\'i}sica Te{\'o}rica and IFIC, Universidad de Valencia-CSIC,
Dr. Moliner 50, 46100-Burjassot, Spain}

\author{Diogo O. Soares-Pinto}

\affiliation{Instituto de F\'{i}sica de S\~{a}o Carlos, Universidade de S\~{a}o Paulo,
CP 369, 13560-970, S\~{a}o Carlos, SP, Brazil}

\author{S\'{i}lvio M. \surname{Duarte Queir\'{o}s}}

\affiliation{Centro Brasileiro de Pesquisas F\'{i}sicas and National Institute of Science and Technology for Complex Systems, 150 Rua Dr. Xavier Sigaud, 22290-180 Rio de Janeiro, Rio de Janeiro, Brazil}

\begin{abstract}
We introduce an analytically treatable discrete time quantum walk in a one-dimensional lattice which combines non-Markovianity and hyperballistic diffusion associated with a Gaussian whose variance, $\sigma _t ^2$, growing cubicly with time, $\sigma \propto t^3$. These properties have have been numerically found in several systems, namely tight-binding lattice models. For its rules, our model can be understood as the quantum version of the classical non-Markovian `elephant random walk'  process, for which the quantum coin operator only changes the value of the diffusion constant though, contrarily to the classical coin.
\end{abstract}

\maketitle

\paragraph{Introduction.}

The random walk problem %--- or the continuous time Brownian motion  --- 
has been a cornerstone in the classical description of systems for which a deterministic approach is either impossible or too complex to be carried out in an efficient way. Equilibrium and non-equilibrium problems like Hamiltonian Monte Carlo, belief propagation, genetic and search algorithms or pricing financial derivatives \cite{binder1986introduction,fishman2013monte, hoos2004stochastic,page2001method,goldberggenetic} are systematically understood as a random walk in phase-space of the respective system. The first fundamental property of a random walk process, \mbox{$X=\left\{ X_t \right\}$}, concerns the time dependence of the variance, $\sigma^2 _t \propto t$. Second, because it derives from a Bernoulli process, the random walk, abides by the ubiquitous Markovian property~\cite{schmidt1985cw},  according to which a memoryless random process is defined as a orderly succession of events where the conditional probability distribution of the future state  $X_t$ (discrete time \mbox{$t> t_0$}) does only depend on its present state,
%
%\begin{equation}
\mbox{$P\left( X_t \, | \, X_{t-1}, \ldots, X_{t_0} \right) = P\left( X_t \, | \, X_{t-1}\right)$}.
%\end{equation}
%

While in the classical treatment of a Physical system probability is above all a tool for getting quantitative answers, in quantum theory, probability is intrinsic \cite{van1983stochastic} and thus quantum walks  emerged as the formal quantum equivalent to random walks~\cite{Davidovich1993, venegas2012quantum}. Physically, quantum walks  describe situations where a quantum particle is moving on a discrete grid, which allows simulating a wide range of transport phenomena \cite{mohseni2008environment,debbasch2012propagation, di2013quantum, di2016quantum, marquez2017fermion} including the description of some types of topological insulators and yields an important approach in quantum computing processes~\cite{childs2009universal,childs2004spatial, portugal2013quantum} . In other words, the particle dynamically explores a large Hilbert space, $\mathcal{H}_P$, spanned by its positions on a lattice corresponding to basis states $\left\{\left| l \right\rangle \right\}$, ($l$ $\in$ $\mathbb{Z}$), that is augmented by a Hilbert space, $\mathcal{H}_C$, spanned by the particle internal states -- \textit{e.g.} a two-dimensional basis $\left\{\left| \uparrow \right\rangle , \left| \downarrow \right\rangle \right\}$.
The evolution of a quantum walk on the full Hilbert space, $\mathcal{H} \equiv \mathcal{H}_{C} \otimes \mathcal{H}_{P}$, is ruled by the combined application of two unitary operators
\begin{equation}
\hat{\mathcal{U}}=\widehat{\mathcal{S}}.\left[\hat{\mathcal{C}}\otimes\hat{\mathcal{I}}\right],
\label{eq:evolop}
\end{equation}
where $\hat{\mathcal{I}}$ is the identity operator on the $\mathcal{H}_P$ subspace. Bearing in mind the analogy of quantum walks  with the classical random walk, the operator $\hat{\mathcal{C}}$ acts on subspace $\mathcal{H}_C$ and plays the same role as the coin. For that reason, it is named \emph{quantum coin} and the internal states related to the subspace $\mathcal{H}_C$ the \emph{coin states}. On the other hand, the \emph{shift operator}, $\widehat{\mathcal{S}}$, is state-dependent and following Ref.~\cite{Davidovich1993} reads
\begin{equation}
\hat{\mathcal{S}}= \sum_l \left| l+1\right\rangle \left\langle l \right| \otimes \left| \uparrow \right\rangle \left\langle \uparrow \right| +
\sum_l \left| l-1\right\rangle \left\langle l \right| \otimes \left| \downarrow \right\rangle \left\langle \downarrow \right| .
\end{equation}
Assuming the quantum coin as
\begin{equation}
\hat{\mathcal{C}} =  \begin{pmatrix} \cos\theta & i \sin\theta \\  i \sin\theta  & \cos\theta \end{pmatrix},
\end{equation}
the successive application of the time-evolution operator, $\hat{\mathcal{U}}$, $t$ times to the initial state
\begin{equation}
\left| \Psi \right\rangle _{0} = \left| l_0 \right\rangle  \otimes \left| s \right\rangle
\equiv \begin{pmatrix} \psi ^\uparrow _0 (l) \\ \psi ^\downarrow _0 (l) \end{pmatrix},
\label{eq:psi0-def}
\end{equation} 
where the internal state is defined as
\begin{equation}
\hspace{-0.25cm}
\left| s \right\rangle \equiv \cos \left(\frac{\gamma}{2}\right) \, \left| \uparrow \right\rangle + \exp^{-i \phi} \, \sin\left(\frac{\gamma}{2}\right) \, \left| \downarrow \right\rangle =
\begin{pmatrix}\cos \frac{\gamma}{2} \\ \mathrm{e}^{-i \phi}\sin \frac{\gamma}{2} \end{pmatrix},
\end{equation}
allows us to obtain the normalised probability at time $t$, \mbox{$P_t(l)= \mathrm{Tr}_s \left\langle \Psi | \Psi \right\rangle _t $}. A straightforward computation (see e.g.~\cite{Davidovich1993}) shows that $\sigma  _t \propto t^2$; in other words, the standard quantum walks  diffuses ballistically in opposition to the classical random walk.

Although systematically ignored due to the matching of a plethora of theoretical predictions with experimental results, there are many processes for which the Markovian property does not hold and therefore they depend on their past. Classically, history-dependent -- i.e., non-Markovian -- processes are often related to anomalous diffusion where the variance of the stochastic process,  grows as $ t^\alpha$ with the diffusion exponent, $\alpha$, different from unity. Instances of physical and biological systems exhibiting subdiffusion ($0<\alpha <1$) or superdiffusion ($1 < \alpha < 2$) are galore \cite{klages2008anomalous, lacasta2004subdiffusion, lutz2003anomalous,cavagna2013diffusion}. Moreover, in several cases such as random search strategies -- namely foraging -- non-Markovian processes have shown to outperform Markovian proposals \cite{viswanathan2011physics}.

The follow-up of the analogy between random walks and quantum walks  has to do with the Markovian nature of the processes. 
In spite of not having a consensual definition \cite{breuer2012foundations}, non-Markovianity in quantum walks  has been extensively studied because memory effects can be taken as an indicator for the presence of canonical quantum properties. Mainstream examples thereof are $XX$-Heisenberg spin chains in a transverse magnetic field $h$~\cite{apollaro2011memory}, Bose-Einstein condensate systems with impurities \cite{haikka2011comparing} and  transport properties of particles in a quenched random media showing Anderson localization, which can interpreted as a memory feature; in other words, when the particle is coupled to a disorded system, it can `remember' and localizes near its initial position. Different regimes, ranging from a canonical ballistic to subdiffusion (delocalization), has been already studied \cite{kitagawa2010exploring, bergmann1984weak} for several kind of randomness and memory. Moreover, non-Markovianity can be explored by means of the history-dependence of the paths \cite{rohde2012quantum, rohde2012entanglement}
Alternatively, in~\cite{hufnagel2001superballistic}, it was first introduced the possibility of measuring hyperballistic diffusion in 1D (quasi-) lattices, a phenomenon later verified on tight-binding lattice models \cite{zhang2012quantum, gholami2017noise}, XXZ spin chains \cite{fitzsimons2005superballistic}, phononic heat transport \cite{tang2017superballistic} and quantum kicked rotors \cite{qin2014dynamical}, where $\alpha >2$ with $\alpha = 3$ playing a leading role and the remaining cases with $2<\alpha <3$ obtained by assuming internal sub-lattices with standard features (for details please consult \cite{guarneri1989spectral, ketzmerick1997determines}).

To the best of our knowledge, while there exists a series of systems exhibiting either non-Markovianity or numerically found hyperballistic behavior, there is not a model that shows both the former and the latter features, namely in an analytical way. If we continue resorting to the analogy between quantum  and random walks, it is not odd to reckon that quantum hyperballistic diffusion would play a similar qualitative role to classical superdiffusion, a fact that emphasises the relevance of introducing a model with such traits. In this letter, we fill the aforementioned gap by considering a quantum walk version of the so-called elephant random walk ~\cite{schutz2004elephants}, one of the few (simple) classical cases where microscopic dynamical rules are translated into non-Markovian statistical properties of the walker.

\paragraph{The classical elephant.}

According to Sh\"{u}tz and Trimper~\cite{schutz2004elephants}, the elephant random walk describes displacements on the infinite discrete lattice $X_t \in \mathbb{Z}$ assuming discrete time as well. The `elephant' starts its walk at some specific point $X_0$ at time $t = 0$ and for $t>t_0$ the stochastic evolution, \mbox{$X_{t} = X_{t-1} + \Delta _{t}$}, occurs as follows: \medskip \\
\emph{(A)} For $t=1$, the `elephant'  moves to the right ($\Delta_1 = 1$) with probability $q$ and to the left ($\Delta_1 = -1$) with probability $1-q$;\medskip \\
\emph{(B)} for $t>1$, an instant in the past $t^\prime < t$ is first randomly and independently chosen abiding by a uniform probability and then 
$\Delta_{j}$ is determined by the rule:  $\Delta_{j} = \Delta_{j^\prime}$ with probability $p$ and $\Delta_{j} = -\Delta_{j^\prime}$ with probability $1-p$.
It is after this rule the particle in this process is dubbed `elephant': it will remember all its previous states.
Accordingly, the conditioned probability distribution of the classical walker displacement at time $t$ was calculated in Ref.~\cite{schutz2004elephants} and reads
\begin{equation}
P\left(\Delta_{t+1}=\ell|\Delta_{t},\ldots,\Delta_{1}\right) = 
\sum_{j=1}^{t}\frac{1-\left(1-2p\right)\ell\,\Delta_{j}}{2 \, t},
\label{eq:probC}
\end{equation}
wherefrom the conditioned moments of the displacement can be computed. Thus, it was demonstrated that the memory parameter $p$ governs the long-term behavior of such process: for $p <1/2$, the `elephant' is a Brownian walker, $\mu _t \equiv  \left\langle X_t \right\rangle \sim X_0$ and $\sigma ^2_t \propto t$; for $1/2<p<3/4$ it becomes a biased Brownian walker with $\mu _t \propto t^{2p-1}$ whereas for $p>3/4$ besides the bias behaviour, the motion becomes superdiffusive with $\sigma ^2_t \propto t^{4p-2}$. Yet, it can be shown that the distribution is  Gaussian and the continuous time limit yields a Fokker-Planck Equation equivalent to that of a Brownian walker subjected to a time-dependent drift force $f_t\left(X \right)= \left(2p-1\right) \left(X_t - X_0 \right)/t$. 

\paragraph{The quantum elephant.}

To formalise the quantum analogue of the elephant random walk, we extend the definition of the quantum walk to a time dependent shift in position 
\begin{eqnarray}
\widehat{\mathcal{S}}^E _{t+1} &= &\frac{1}{t}\sum_{j=\text{1}}^{t} \left\{ \sum _l \left| l+\Delta_{j} \right\rangle _{t+1} \left\langle l \right| _t \right. \otimes \left| \uparrow \right\rangle \left\langle \uparrow \right| \nonumber \\ 
& +& \left. \sum _l  \left| l-\Delta_{j}\right\rangle  _{t+1} \left\langle l \right| _t
\otimes \left| \downarrow \right\rangle \left\langle \downarrow \right|
\right\} , \quad \left(t\geq1\right). 
\label{eq:shift1}
\end{eqnarray}
As the quantum version of rule $(A)$, we consider a set of random variables $\Delta _{j}$, uniformly distributed at each time step of the quantum walk evolution. The quantum equivalent to rule $(B)$ is to consider that at time $t+1$, the amplitude $\psi^\uparrow _{t+1}(l +\Delta _{j})$ encodes the probability to move towards the right by a step of size $\Delta _{j}$ and  the amplitude $\psi^\downarrow _{t+1}(l -\Delta _{j})$ encodes the probability to move towards the left by a step of size $\Delta _{j}$.

\begin{figure}[h]
\includegraphics[width=0.49\columnwidth]{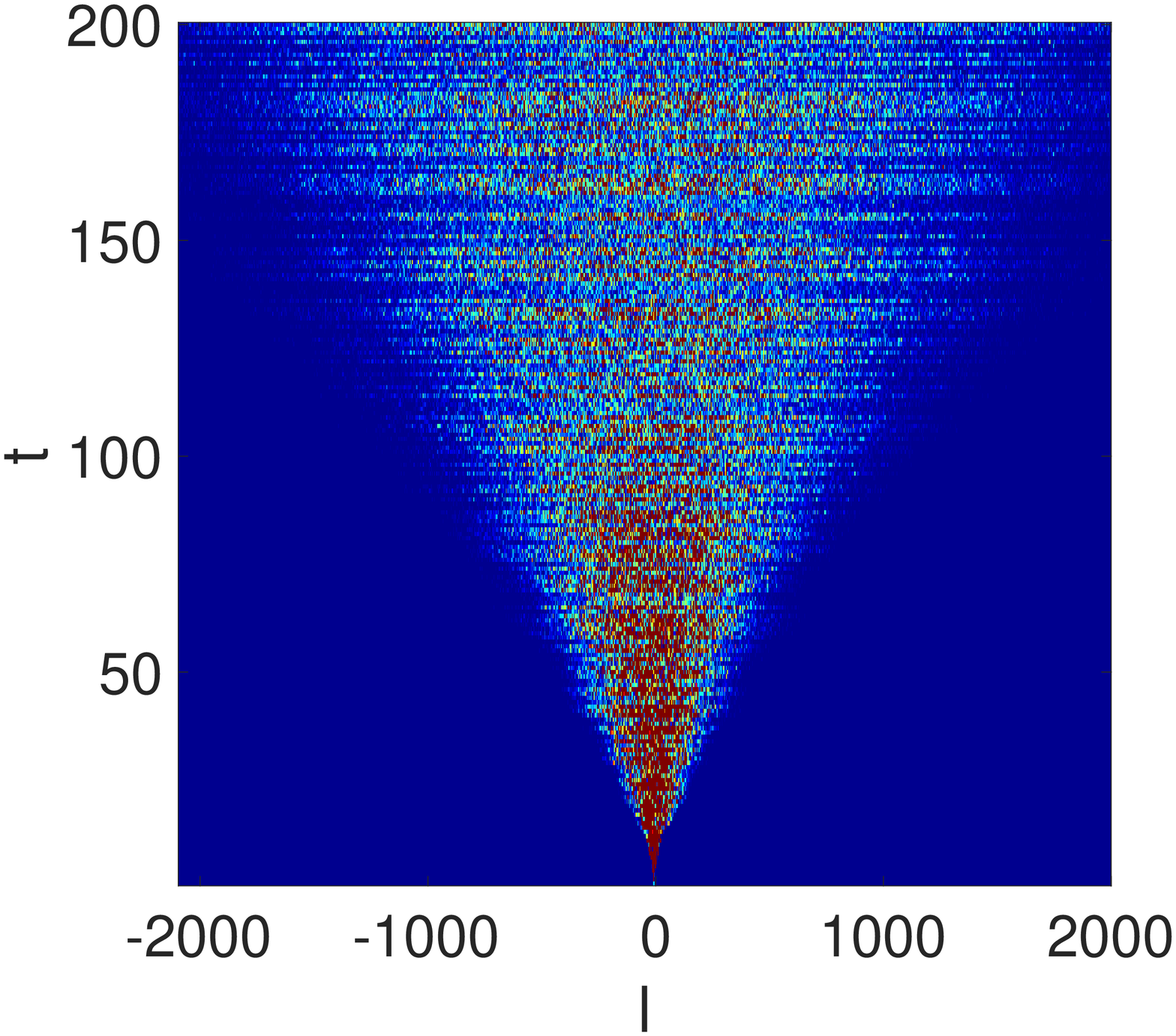}
\includegraphics[width=0.49\columnwidth]{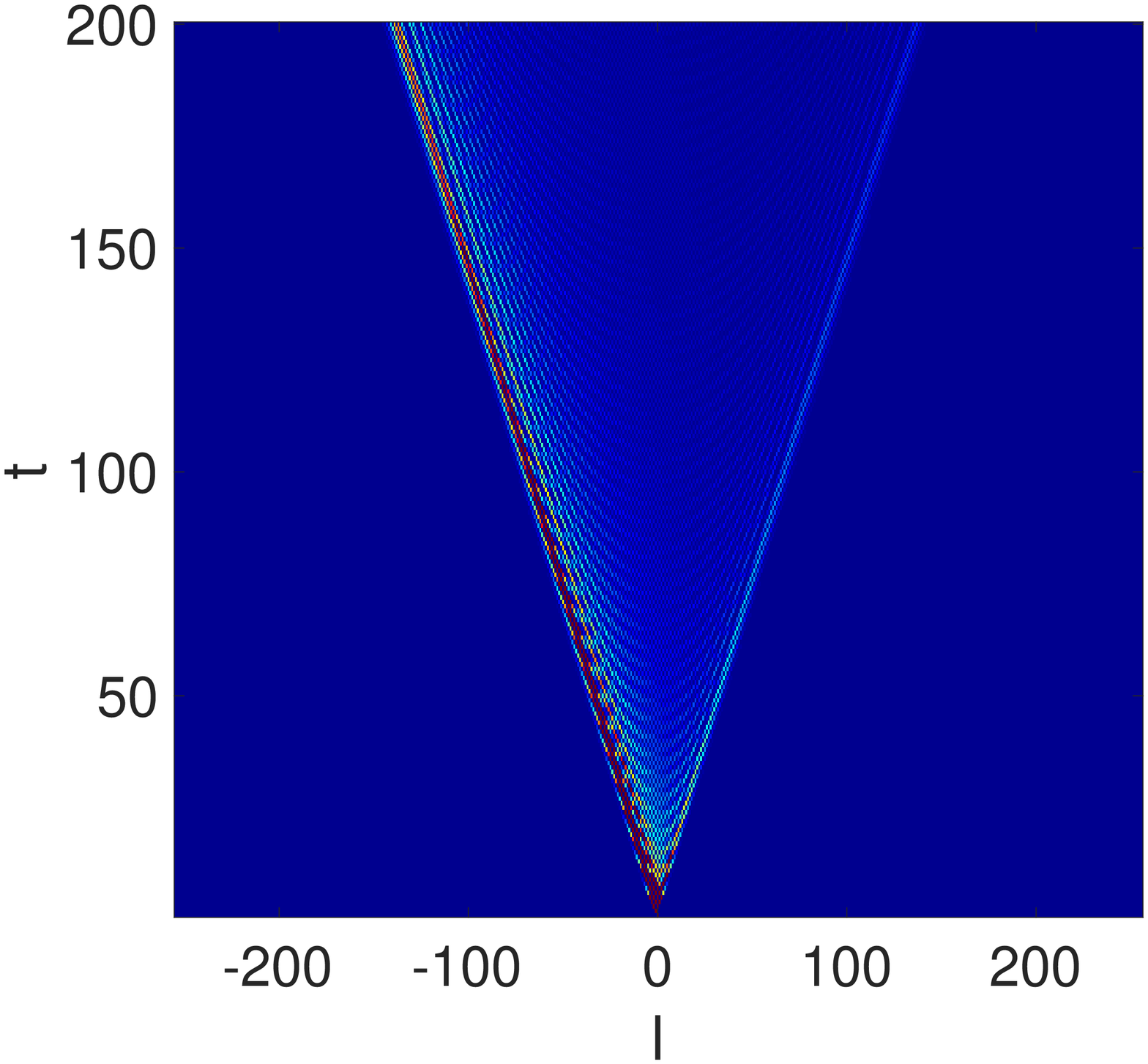}
\includegraphics[width=0.49\columnwidth]{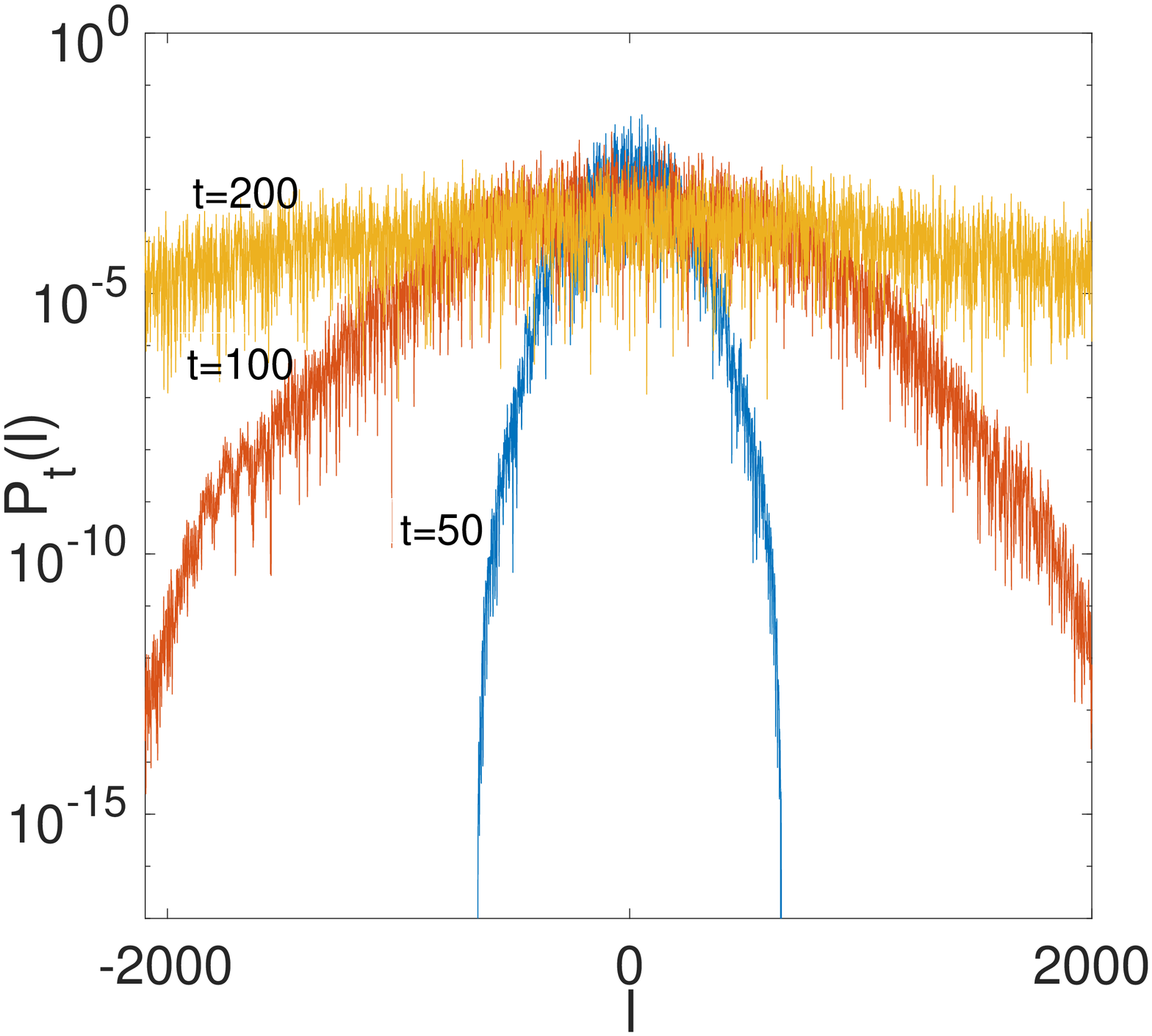}
\includegraphics[width=0.49\columnwidth]{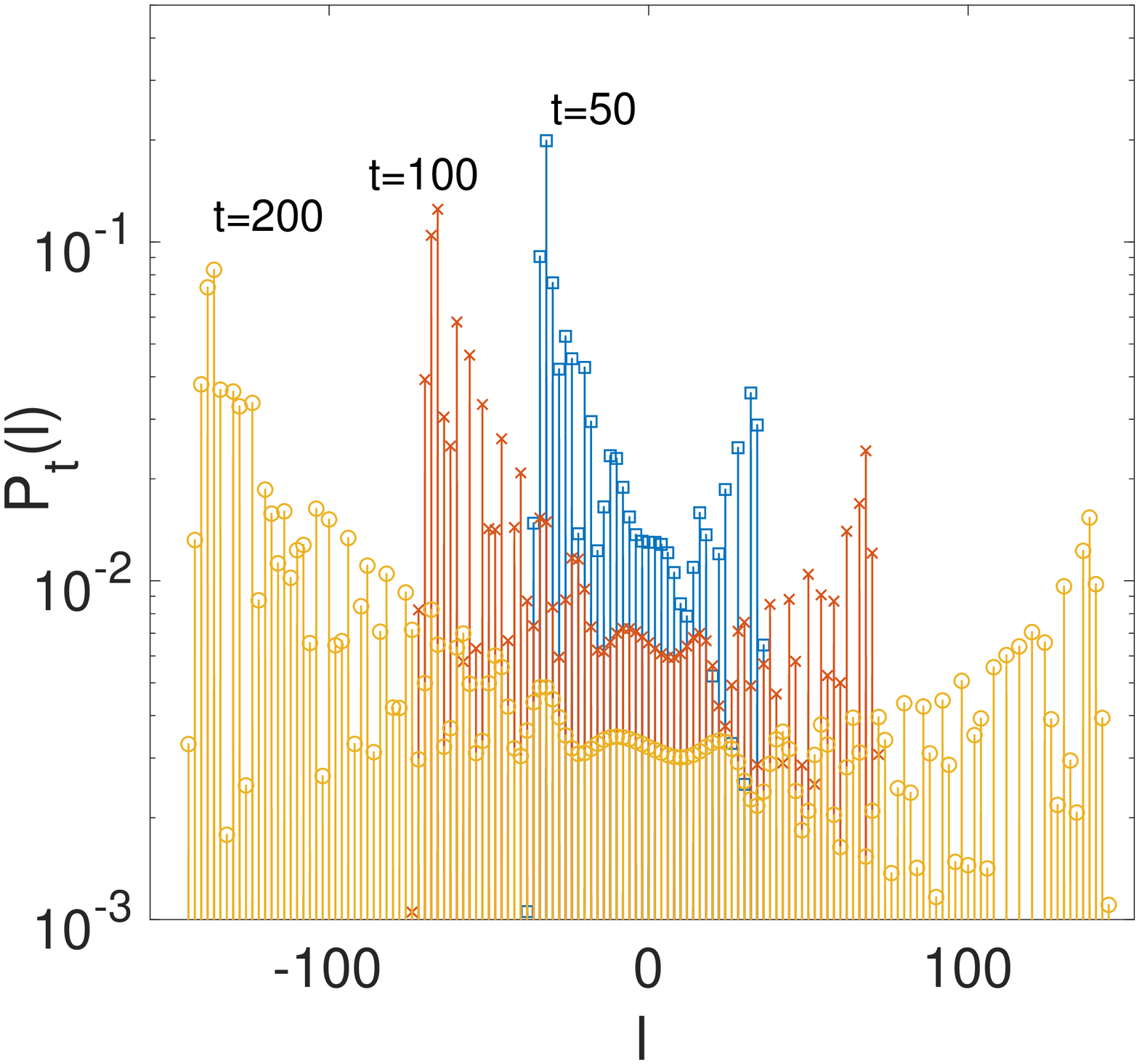}
\caption{(Color online) The panels on the left are for the elephant quantum walk and the panel on the right for the quantum walk for $\theta =\frac{\pi}{4}$. (Top) Time evolution of the probability distribution, whose cones are  superlinear and linear, respectively. The initial state is localized at the origin, with a coin state $|s\rangle = \frac{1}{\sqrt{2}}(1,1)^{T}$. (Bottom) Probability distribution at different time steps.}
\label{fig:diff}
\end{figure}

Having said that, the dynamical evolution defined by Eq.~\eqref{eq:evolop} for the elephant random walk is replaced by
\begin{equation}\widehat{\mathcal{U}}_{t}=\widehat{\mathcal{S}}^{E}_t .\left[\hat{\mathcal{C}}\otimes\hat{\mathcal{I}}\right].
\label{eq:newevolop}
\end{equation}
Therefore, a significant difference immediately emerges: the evolution operator ceases being the simple iteration of the one-step operator, i.e., $\hat{\mathcal{U}}_{t}\neq\left[\hat{\mathcal{U}}_{1}\right]^{t}$, a property that hints at the non-Markovian nature of the  elephant quantum walk, which we shall prove shortly. Actually, the equality, $\hat{\mathcal{U}}_{t} = \left[\hat{\mathcal{U}}_{1}\right]^{t}$, can be understood as the quantum analogue of the classical condition that the transition matrix $\mathsf{T}$ of a Markov chain after $t$ steps yields, $\mathsf{T}_{t}=\mathsf{T}_{1}^{t}$. Moreover, a direct calculation, allows us to introduce the conditioned probability density functions for the jumps in the elephant quantum walk similarly to Eq.~\eqref{eq:probC}. For $t = 2$ that reads
\begin{widetext}
\begin{eqnarray}
P\left(\Delta_{2}|\Delta_{1}\right)=\left[\cos\left(\frac{1-\Delta_{2}}{4}\pi-\theta\right)\right]^{2}\left\{ 1+\Delta_{1}\left[\cos\left(\gamma\right)\cos\left(2\theta\right)+\sin\left(\gamma\right)\sin\left(2\theta\right)\sin\left(\phi\right)\right]\right\}
\end{eqnarray}
and for $t\geq2$,
\begin{eqnarray}
P\left(\Delta_{t+1}=\Delta |\Delta_{t},\Delta_{t-1},\ldots,\Delta_{1}\right) =  \frac{\left(t-1\right)!}{t!}\sum_{j=1}^{t}\prod_{k=2}^{t}\left[\cos\left(\frac{\Delta_{k+1}-\Delta_{k}}{4}\pi-\theta\right)\right]^{2}P\left(\Delta_{2}|\Delta_{1}\right)\,\delta_{\Delta,\Delta_{j}} \\
 +\frac{1}{t!}\sum_{j=1}^{t}\left[A\left(t,\left\{ \Delta_{t}\right\} \right)\left[\cos\theta\right]^{2\left(t-2\right)}+B\left(t,\left\{ \Delta_{t}\right\} \right)\left[\sin\theta\right]^{2\left(t-2\right)}\right]\,P\left(\Delta_{2}|\Delta_{1}\right)\,\delta_{\Delta,-\Delta_{j}},\nonumber 
\end{eqnarray}
\end{widetext}
where the coefficients $A\left(t,\left\{ \Delta_{t}\right\} \right)$ and $B\left(t,\left\{ \Delta_{t}\right\} \right)$ represent the number of sequences yielding an even(odd) number of contrarian steps composing the sequence so that the condition $A\left(t,\left\{ \Delta_{t}\right\} \right)+B\left(t,\left\{ \Delta_{t}\right\} \right)=\left(t-1\right)!$ is verified. Notice that for $t\geq2$, and due to the quantum nature of this process, we cannot have the conditioned probability abiding by a simple superposition of contributions involving the gauging value $\Delta_{t+1}=\ell$ and a past chosen step. Each term involves all the values of the chain. This contains a signature of a long-range (non-Markovian) memory effect on the walker dynamics, giving rise to a completely different behavior from the usual quantum walks , as we can see in Fig. \ref{fig:diff}.

Regardless of the different definitions of Markovianity in quantum walks ~\cite{rivas2014quantum, breuer2016colloquium}, we assess the non-Markovianity of the elephant quantum walk following \cite{breuer2009measure, laine2010measure} where it is shown that if the discrete analogous of the trace distance velocity, $v_t \equiv D_{t+1} - D_t$\footnote{$D_t \equiv \frac{1}{2}|\omega^A_t -\omega^B_t |$, where $\omega  ^{(\cdot)} _j$ represents the trace of the reduced density matrix operator at time $t$ given the initial condition $(\cdot)$, defined as $ \hat \rho^c _t =\sum_l \langle l | \Psi_t \rangle \langle \Psi_t | l \rangle$, for the pure states defined in Eq. \ref{eq:psi0-def}.} , is positive at least once, then the process is non-Markovian.

\begin{figure}[h]
\centering
\includegraphics[width=0.49\columnwidth]{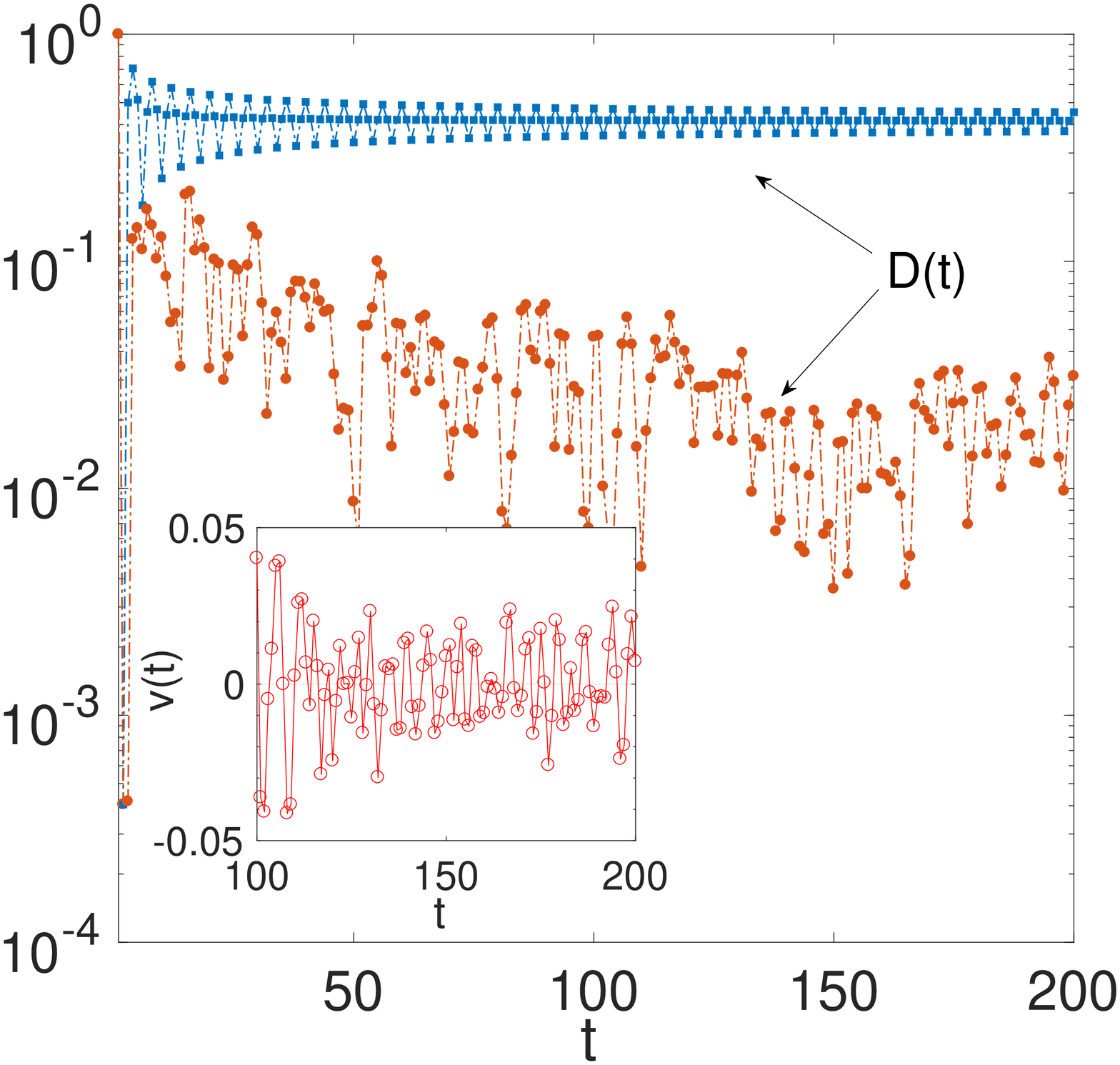}
\includegraphics[width=0.49\columnwidth]{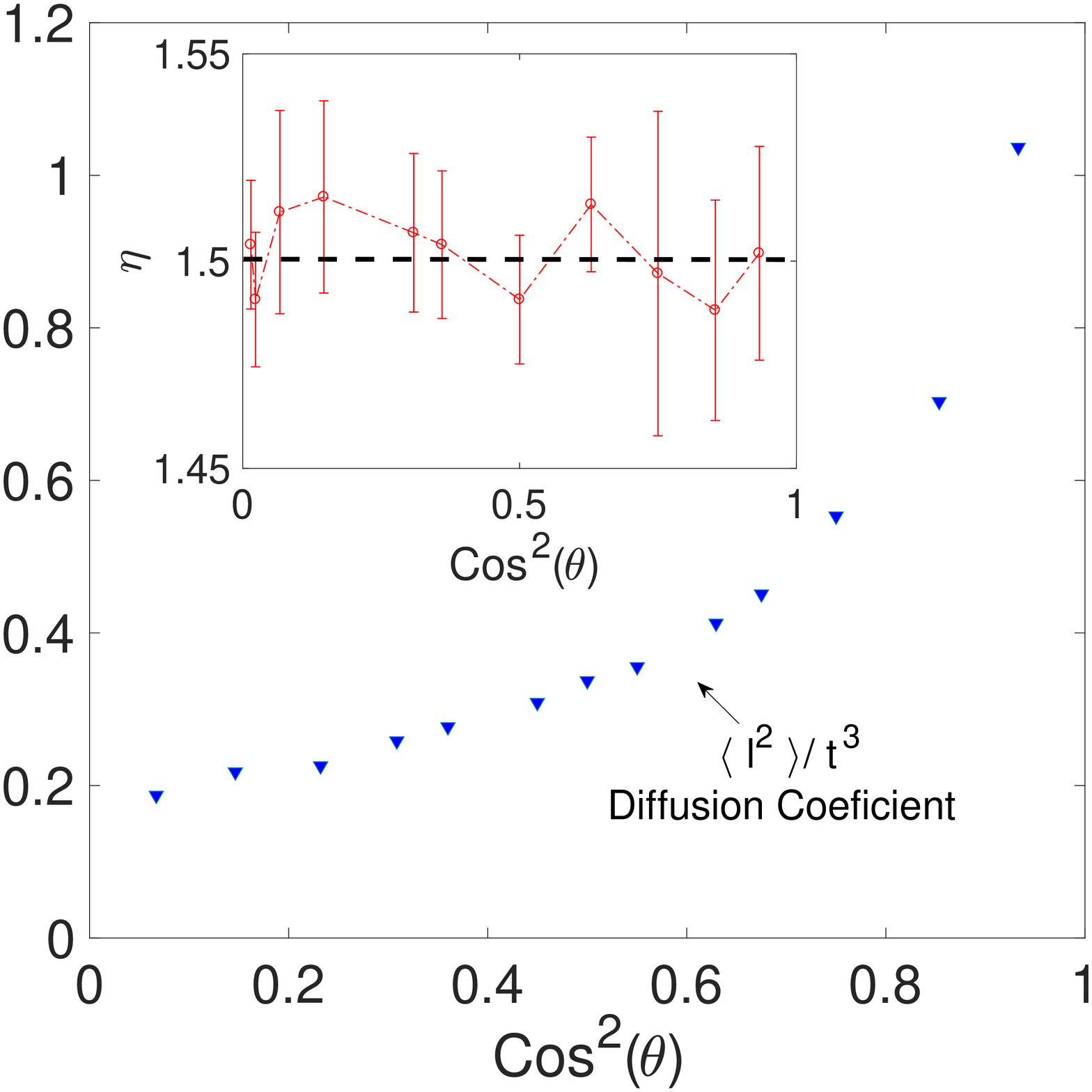}
\caption{(Color online) Left: the squares (blue) represents the trace distance, $D(t)$,   {\it v.} the number of time steps between two initial states with $\gamma _A=0$ and  $\gamma _B=\pi$, $\phi=0$, for an initial Gaussian packet of width $\delta = 0.001$. In case of noise, the trace distance decrease rapidly after the second time step (circles, red). The inset shows the velocity of the trace distance. Right: Diffusion coefficient --- and in the inset the exponent $\eta = \alpha / 2$ ---, {\it v.} $\theta$, for the same initial condition.}
\label{fig:trace}
\end{figure}

In Fig.~\ref{fig:trace}, we present the computation of the trace distance and its velocity for two initial pure states representing opposite poles on the Bloch sphere (north and south) with $\gamma _A=0$ and  $\gamma _B=\pi$, $\phi=0$, for an initial Gaussian packet of width $\delta = 0.001$ that is compared with the standard quantum walk employing the Hadamard coin. The difference between both is notorious \cite{hinarejos2014chirality} and the trace distance of the reduced density operator $ \hat \rho^c _t $ displays typically a non-monotonic behavior in free-decoherence cases. In presence of decoherence, the trace distance goes asymptotically to zero and the process becomes fully Markovian. In our case, surprisingly, the trace distance velocity, $v_t$ is positive more than once and the trace distance is non zero for a long period of the elephant quantum walk's life, viz., displaying non-Markovianity, still in presence of dynamical noise. 

After shedding light on the non-Markovianity of the elephant quantum walk, we focus on the probability distribution  and diffusion properties;  following the same techniques introduced by \cite{Brun2003a,Annabestani2009}, we were able to analytically prove that our process is associated with a Gaussian distribution,
\begin{equation}
P_t(l) = \exp \left[ -l^2 / \left( 2\, \sigma ^2 _t \right) \right] / \sqrt{ 2\, \pi \, \sigma ^2 _t} ,
\end{equation}
where the variance reads
\begin{equation}
\langle \ell ^2 \rangle = \int^{\ell =\infty}_{\ell =-\infty} \ell ^2 \langle r(t) \rangle d\ell  = 2\sqrt{2 \pi} ( C_1 + 2 C_2 ) t^3,
\end{equation}
with $C_1, C_2$ being real coefficients depending on the coin parameter $\theta$. The analytical details are fully presented in the Appendix. Explicitly, the elephant quantum walk is both Gaussian and robustly hyper-ballistic, features that have been numerically found in other models as previously cited. These results have been confirmed by the dynamical implementation of the walk as depicted in Fig. \ref{fig:trace} (right panel). 

\paragraph{Concluding Remarks.}

In this work, we have established the first analytically treatable model exhibiting both hyper-ballistic diffusion and non-Markovianity, which can be interpreted as the quantum version of the so-called `elephant random walk'. Our model yields an exact diffusion exponent equal to $3$, a value that has been numerically found in a variety of systems ranging from quantum diffusion on tight-binding lattice models to phononic heat transport among other systems we previously cited. At the same, non-Markovianity in quantum walks  has been intensively studied and found in systems as well, but the computation of analytical properties is seldom. In this case, we have been able compute that the quantum elephant has its motion associated a Gaussian, as numerically shown for the model in \cite{rohde2012quantum}.

That said, and bearing in mind its plainness, our elephant quantum walk can be regarded as a dynamical representation of hyper-ballistic non-Markovian such systems so that they can be studied in a more direct way, with a direct connection between Hamiltonian and quantum coin parameters, as happens with Hamiltonian regular maps in non-linear dynamics \cite{tabor1989chaos}. Accordingly, the mapping of the parameters of systems yielding $\alpha = 3$ and ours are clearly worth of carrying out.

From a quantum implementation perspective, apparatus equivalent to those considering cold atoms in optical lattices \cite{stutzer2013superballistic, billy2008direct, roati2008anderson}, which manage to yield super-ballistic diffusion are straightforward options to performing elephant quantum walks . Alternatively, we can understand the introduction of such memory as the storage process on the state of the system that is cyclically isolated and put in contact with a surrounding environment which acts as the tossing of the coin. In practical terms, this can be carried out considering an apparatus close to that present in Ref. \cite{liu2011experimental} that accommodates that long-time storage.

Since the systems studied in Refs. \cite{hufnagel2001superballistic, zhang2012quantum} can also show super-ballistic diffusion with an exponent less than $3$, in future work we will loose the form of the memory by assuming a kernel similar to Ref. \cite{queiros2007generalised} that is able to represent all the cases spanning from white-noise to uniform dependence passing through exponential and power-law.

Last, taking into account that classical super-diffusive search approaches are robustly more efficient than Brownian searches, we consider it is worth looking at the elephant quantum walk within the context of search algorithms as an (improved) extension of the random walk search algorithm \cite{shenvi2003quantum, chakraborty2016spatial}, which fits in the Grover class.

\paragraph{Acknowledgment.}

The authors would like to acknowledge the financial support from Brazilian funding agencies CNPq, FAPESP, the Brazilian National Institute of Science and Technology of Quantum Information (INCT/IQ), FAPERJ and the Spanish Ministerio de Educación e Innovación (MICIN-FEDER project) and Generalitat Valenciana grant and the project PEPS "QNet" from the CNRS. 

\appendix

\section{Illustration of the random and quantum models}

\begin{figure}[h]
\centering
\includegraphics[width=0.8\columnwidth]{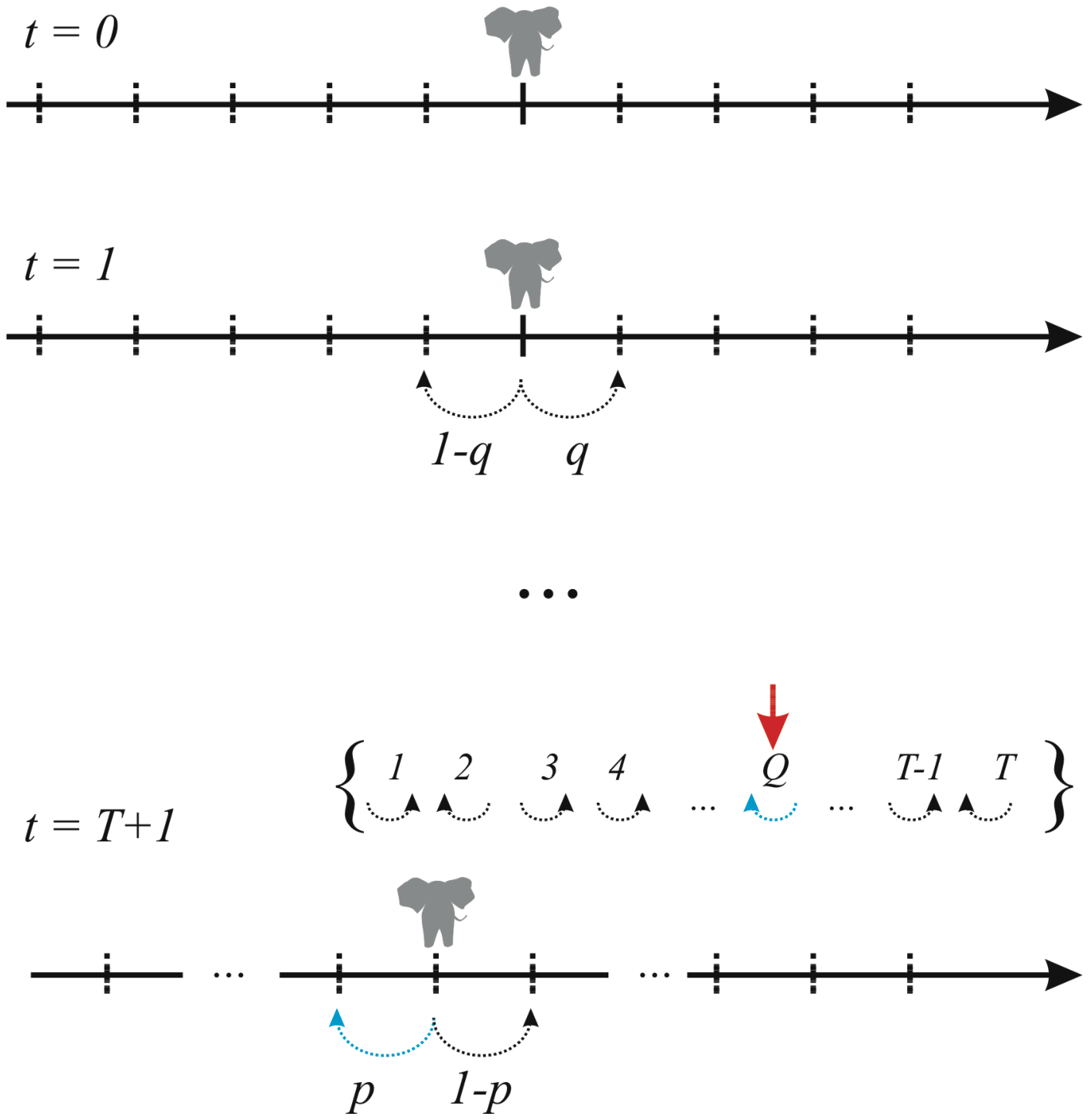}
\includegraphics[width=0.8\columnwidth]{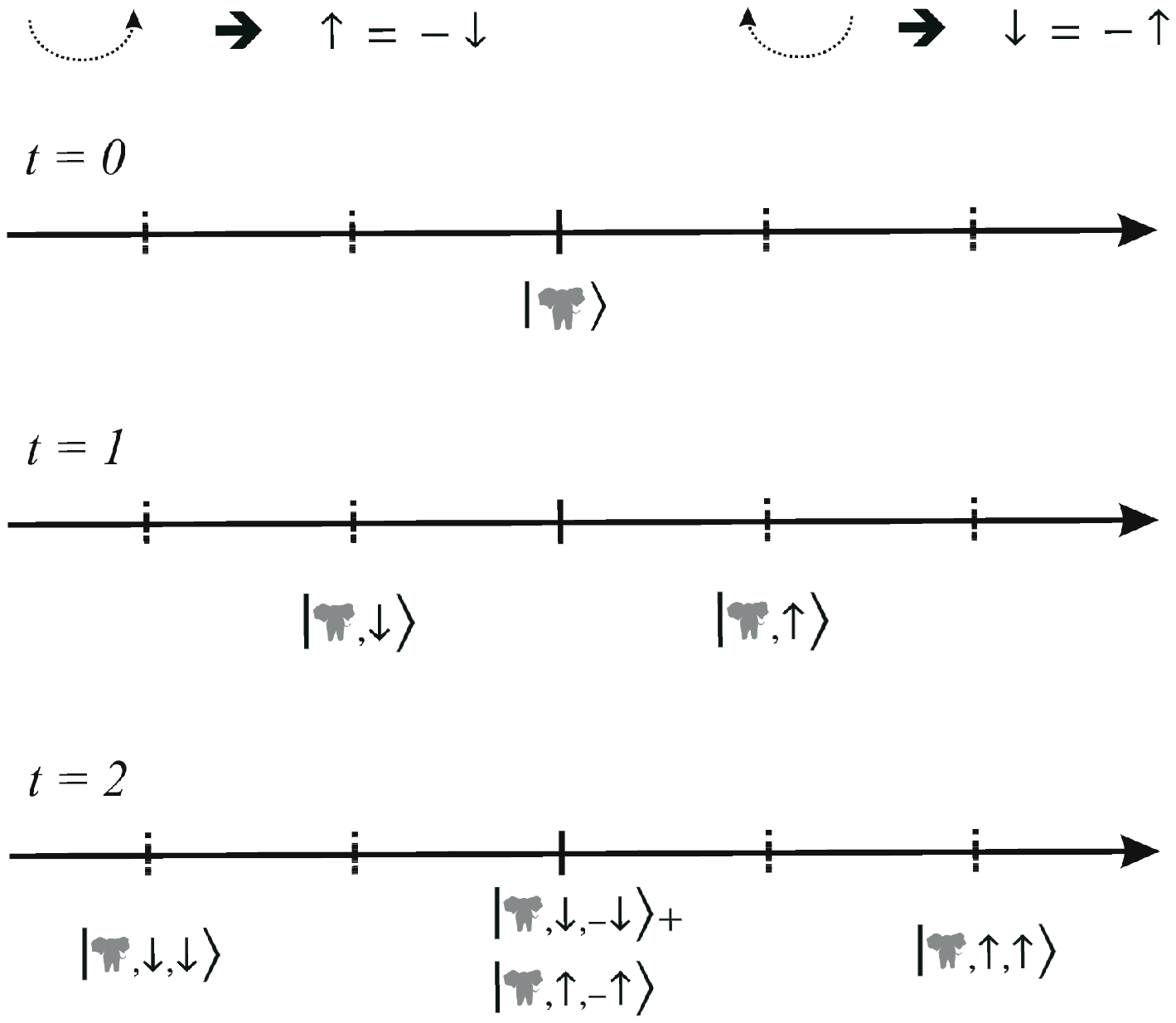}
\caption{Top: Depiction of the elephant random walk model. Down: Depiction of the elephant quantum walk model.}
\label{fig:algos}
\end{figure}

In Fig.~\ref{fig:algos}, we present an illustration of the random and quantum elephant approaches.
At time $t=1$, the elephant moves either to the right or to the left with probability $q$ and $1-q$, respectively, Then, at $t=2$, the elephant will take the previous set of displacements --- which is only the $ t = 1$ case --- and decide whether it will repeat that displacement with probability $p$ or do the opposite  with probability $1-p$. Generically, at time $t = T + 1$, the elephant selects (uniformly) one of the previous instants --- $t = Q$ in the case of the illustration --- and determines to do the same or not afterwards.

For the quantum case, the rules are basically the same, i.e., at each time step $T+1$, we have a collection of $T$ displacements that are the outcome of the interaction between the coin and the spin. At time $t =1$, the evolution operator has into account a possible motion to the left and a possible motion to the right. For $t =2$, --- which defines the state at $t = 3$ --- the operator has to take into account the two states of the spin that can be assumed at $t = 1$ and the fact that at $t = 2$ one can have the previous state repeated or flipped so that at a given time $T+1$ the evolution operator corresponds to a equally weighted superposition --- and then the uniform distribution --- of each of the $T$ previous operators plus the opposite situation, conjugated with the operator at that time, as indicated by Eq.~\eqref{eq:shift1} in the main text.

\section{Gaussianity and Diffusion Exponent}

Regarding the statistical properties of this process, following the techniques introduced by \cite{Brun2003a,Annabestani2009}, we can transform the density matrix into a four dimensional column vector in order to use the affine map approach yielding:
\begin{equation}
\hat \rho ^c_t \equiv\frac{1}{2}\left[r ^{\left( 0 \right)}_t \hat {\mathbf{I}}+r^{\left( 1 \right)}_t \hat {\boldsymbol{\sigma}}_{1}+r^{\left( 2 \right)}_t \hat {\boldsymbol{\sigma}}_{2}+r^{\left( 3 \right)}_t \hat {\boldsymbol{\sigma}}_{3}\right],
\end{equation}
where $\hat {\boldsymbol{\sigma}}_{i}$ with $i=\{1,2,3 \}$ are the Pauli matrices, and $r^{\left(i\right)}_t=\mathrm{Tr}\, \{\hat \rho ^c_t \,\hat {\boldsymbol{\sigma}}_{i}\},$ from which we define the vector $\mathbf{r}_t=\left(r^{\left(i\right)}_t \right)^{T}$ with $r ^{\left( 0 \right)}_t =1$. The evolution of this vector reads
\begin{equation}
\mathbf{r}_t=\int_{-\pi}^{\pi}\frac{dk}{2\pi}L_{k,t}^{t}r(0),
\end{equation}
where $L_{k,t}$ is a matrix evolution operator, namely the Limbladian,
\begin{equation}
L_{k,t}=\left(\begin{array}{cccc}
1 & 0 & 0 & 0\\
0 & L_{22,t} & L_{23,t} & L_{24,t}\\
0 & L_{32,t} & L_{33,t} & L_{34,t}\\
0 & L_{42,t} & L_{43,t} & L_{44,t}
\end{array}\right),
\end{equation}
with matrix elements:
\begin{align*}
L_{22,k,t}= & \cos(2 k \Delta_t)\\
L_{23,k,t}= & \cos(2 \theta) \sin(2 k \Delta_t)\\
L_{24,k,t}= & \sin(2 k \Delta_t) \sin(2 \theta)\\
L_{32,k,t}= & -\sin(2 k \Delta_t) \\
L_{33,k,t}= & \cos(2 k \Delta_t)\cos(2 \theta)\\
L_{34,k,t}= & \cos(2 k \Delta_t)\sin(2 \theta)\\
L_{42,k,t}= & 0\\
L_{43,k,t}= & -2\cos(\theta)\sin(\theta)\\
L_{44,k,t}= & \cos(2\theta).
\end{align*}

In this case, $\Delta_t$ is chosen randomly at each time-step with a uniform probability
distribution in the interval $[1-t,1+t]$. The evolution matrix $L_{k,t}$ has to be averaged over the interval $[1-t,1+t]$
as: 
\begin{equation}
\langle L_{k,t}\rangle=\frac{1}{2 t} \int_{1-t}^{1+t} L_{k,t'}dt'.
\end{equation}
Therefore, at each time step, the evolution is fully determined by:
\begin{equation}
\langle r_t\rangle=\int_{-\pi}^{\pi}\frac{dk}{2\pi} \langle L_{k,t}\rangle\langle L_{k,t-1}\rangle...\langle L_{k,1}\rangle r_0.
\end{equation}
It can be demonstrated that in the limit $k  \rightarrow 0$, the eigenvalues of $\langle L_{k}\rangle$ are:
\begin{eqnarray} 
\lambda_{1}=1,\\
\lambda_{2} \propto \exp(B_1(\theta) + i C_1(\theta) O(k^2) t^2),\\
\lambda_{3} \propto \exp(B_2(\theta) + i C_2(\theta) O(k^2) t^2)
\\
\text{and} \  \lambda_{4}=\lambda_{3}^{*},
\end{eqnarray}
where $(B_1,B_2, C_1, C_2) $ are positive real depending on the microscopic parameter $\theta$.

Now, exploiting the stationary phase theorem, we can neglect terms like $e^{i \omega_i t}$, with $\omega_i = - i \log(\lambda_i)$, $i=1,2,3$ ,
when time goes to infinity, 
\begin{equation}
\lim_{t\rightarrow\infty}\langle L_{k,t}\rangle\langle L_{k,t-1}\rangle...\langle L_{k,1}\rangle =\left(\begin{array}{c}
1\\
e^{- i C_1(\theta) t^3 k^2}\\
e^{- i C_2(\theta) t^3 k^2}\\
e^{  i C^*_2(\theta) t^3 k^2}
\end{array}\right),
\end{equation}
and in physical space, the above vector finally reads:
\begin{equation}
\langle r_t \rangle =\left(\begin{array}{c}
2 \pi \delta_\ell\\
\frac{\exp(-\frac{\ell^2}{4 C_1 t^3})}{\sqrt{2 C_1 t^3 }}\\
\frac{\exp(-\frac{\ell^2}{4 C_2 t^3})}{\sqrt{2 C_2 t^3 }}\\
\frac{\exp(-\frac{\ell^2}{4 C_2 t^3})}{\sqrt{2 C_2 t^3 }}\\
\end{array}\right).
\end{equation}
whence we can immediately identify a Gaussian for each row.\footnote{A Dirac delta is the limit distribution of a Gaussian with the variance going to zero.}

From the last result it is possible to compute all of the cumulants of the elephant quantum walk. The first and second moment are now given by:
\begin{equation}
\langle\ell_t\rangle = \int^{\infty}_{\ell=-\infty}\ell\langle r_t \rangle d\ell = 0 ,
\end{equation}
and 
\begin{equation}
\langle \ell_t^2 \rangle = \int^{\infty}_{\ell=-\infty} \ell_t^2 \langle r_t \rangle d\ell = 2\sqrt{2 \pi} ( C_1 + 2 C_2 ) t^3,
\end{equation}
which corroborates the hyper-ballistic diffusion behavior. Since the distribution is Gaussian, all the other cumulants vanish.
\vspace{5cm}

\bibliographystyle{apsrev4-1}
\bibliography{library}

\end{document}

%% file: kkmacros.tex
\usepackage{amsmath,amsfonts,amssymb}

\newcommand{\couic}[1]{}